\documentclass[11pt,a4paper]{emulateapj}
\slugcomment{{\sc Accepted to ApJ:} May 8, 2012}
\usepackage{amsmath}
\usepackage{threeparttable}

\shorttitle{Debris Disks in Tucana-Horologium}
\shortauthors{Donaldson et al.}

\begin{document}

\title{Herschel PACS Observations and Modeling of Debris Disks in the Tucana-Horologium Association}

\author{J. K. Donaldson\altaffilmark{1}, A. Roberge\altaffilmark{2}, C. H. Chen\altaffilmark{3}, J.-C. Augereau\altaffilmark{4}, W. R. F. Dent\altaffilmark{5,}\altaffilmark{6}, C. Eiroa\altaffilmark{7}, A. V. Krivov\altaffilmark{8}, G. S. Mathews\altaffilmark{9}, G. Meeus\altaffilmark{7}, F. M\'{e}nard\altaffilmark{4}, P. Riviere-Marichalar\altaffilmark{10},   G. Sandell\altaffilmark{11} }

\altaffiltext{1}{Department of Astronomy, University of Maryland, College Park, MD 20742; \email{jessd@astro.umd.edu}}
\altaffiltext{2}{Exoplanets and Stellar Astrophysics Laboratory, NASA Goddard Space Flight Center, Code 667, Greenbelt, MD 20771}
\altaffiltext{3}{Space Telescope Science Institute, 3700 San Martin Dr., Baltimore, MD 21218}
\altaffiltext{4}{UJF - Grenoble 1 / CNRS-INSU, Institut de Plan\'{e}tologie et d'Astrophysique de Grenoble (IPAG) UMR 5274, Grenoble, F-38041, France}
\altaffiltext{5}{ALMA, Avda Apoquindo 3846, Piso 19, Edificio Alsacia, Las Condes, Santiago, Chile}
\altaffiltext{6}{European Southern Observatory, Alonso de C\'{o}rdova 3107, Vitacura, Santiago, Chile}
\altaffiltext{7}{Dpt.\ F\'{i}sica Te\'{o}rica, Facultad de Ciencias, Universidad Aut\'{o}noma de Madrid, Cantoblanco, 28049 Madrid, Spain}
\altaffiltext{8}{Astrophysikalishes Institut, Friedrich-Schiller-Universit\"{a}t Jena, Schillerg\"{a}\ss{}chen 2-3, 07745 Jena, Germany}
\altaffiltext{9}{Institute for Astronomy (IfA), University of Hawaii, 2680 Woodlawn Dr., Honolulu, HI 96822}
\altaffiltext{10}{Centro de Astrobiolog\'{i}a – Depto. Astrof\'{i}sica (CSIC-INTA), POB 78, 28691 Villanueva de la Ca\~{n}ada, Spain}
\altaffiltext{11}{SOFIA-USRA, NASA Ames Research Center, Building N232, Rm. 146, Moffett Field, CA 94035}

\begin{abstract}
We present {\it Herschel} PACS photometry of seventeen B- to M-type stars in the 30 Myr-old Tucana-Horologium Association.  This work is part of the {\it Herschel} Open Time Key Programme ``Gas in Protoplanetary Systems'' (GASPS).  Six of the seventeen targets were found to have infrared excesses significantly greater than the expected stellar IR fluxes, including a previously unknown disk around HD30051.  These six debris disks were fitted with single-temperature blackbody models to estimate the temperatures and abundances of the dust in the systems.  For the five stars that show excess emission in the Hershcel PACS photometry and also have {\it Spitzer} IRS spectra, we fit the data with models of optically thin debris disks with realistic grain properties in order to better estimate the disk parameters. The model is determined by a set of six parameters: surface density index, grain size distribution index, minimum and maximum grain sizes, and the inner and outer radii of the disk. The best fitting parameters give us constraints on the geometry of the dust in these systems, as well as lower limits to the total dust masses. The HD105 disk was further constrained by fitting marginally resolved PACS 70$\,\mu$m imaging.
\end{abstract}


\section{Introduction}

Debris disks are the last stage of circumstellar disk evolution, in which the gas from the protoplanetary and transitional disk phases has been dissipated and the dust seen comes from collisions between planetesimals.  In the youngest debris disks, $\lesssim100$ Myrs old, terrestrial planets may still be forming \citep{Kenyon06}.  Giant planets must form before the gas dissipates; their gravitational interactions with planetesimals and dust can leave signatures in debris disks.  
Cold debris disks may be Kuiper-belt analogs, signaling the location and properties of planetesimals remaining in the disk.

To be suitable for life, terrestrial planets in habitable zones must have volatiles such as water brought to their surfaces from beyond the ice line.  Planetesimals in Kuiper Belt-like debris disks may provide this reservoir of volatiles \citep{Lebreton12}.  We do not yet have the capability to detect the planetesimals in these disks, but we can detect the smaller dust grains.  These dust grains are believed to be produced through the collisions of the larger planetesimals, and therefore are likely to have similar compositions to the larger undetected bodies.  The properties and locations of the dust grains in Kuiper Belt analogs can provide clues to the properties of the hidden planetesimals.  Additionally, dust structures in the disk may point to unseen exoplanets \citep[e.g.\ $\beta$ Pic b;][]{Lagrange10}.  

To measure the cold dust in the outer regions of debris disks, we need great sensitivity at far-infrared wavelengths where the thermal emission from the cold dust grains peaks ($\ge 70\,\mu$m).  
The {\it Herschel Space Observatory} \citep{Pilbratt10} provides a unique opportunity for sensitive debris disk surveys.  {\it Herschel}'s PACS instrument \citep{Poglitsch10} is sensitive to the cold dust with a wavelength range of $55-210\,\mu$m. Additionally, {\it Herschel}'s spatial resolution is almost 4 times better than {\it Spitzer} at similar wavelengths, and therefore reducing confusion with background galaxies and interstellar cirrus and making it easier for {\it Herschel} to detect faint, cold debris disks.  

In this paper, we present results of a sensitive {\it Herschel} debris disk survey in the 30-Myr-old Tucana-Horologium Association.  This work is part of the {\it Herschel} Open Time Key Programme ``Gas in Protoplanetary Systems'' \citep[GASPS; Dent et al.\ in prep,][]{Mathews10}. The GASPS survey targets young, nearby star clusters with well determined ages, ranging from 1-30 Myrs.  This range of ages covers the stages of planet formation from giant planet formation \citep[$\sim 1$ Myr;][]{Alibert05} to the late stages of terrestrial planet formation \citep[10-100 Myrs;][]{Kenyon06}.  The targets in each group were chosen to span a range of spectral types from B to M.  

\begin{table*}[ht]
\centering
\caption{Stellar Properties in the 30 Myr-old Tucana-Horologium Association\label{tab:stars}}
\begin{threeparttable}
\begin{tabular}{l c c c c c}
\hline\hline
Star & Right Ascension& Declination & Spectral\tnote{a} & T$_{\ast}$\tnote{b} & Distance\tnote{c} \\
     & (J2000)        & (J2000)     & Type     &  (Kelvin)        & (pc)     \\
\hline
HD2884\tnote{d} & 00:31:32.67 & -62:57:29.58 & B9V & 11250 & $43\pm1$  \\
HD16978 & 02:39:35.36 & -68:16:01.00 & B9V & 10500 & $47\pm1$ \\
HD3003\tnote{d}  &00:32:43.91 & -63:01:53.39 & A0V & 9800 & $46\pm1$ \\
HD224392 & 23:57:35.08 & -64:17:53:64 & A1V & 9400 & $49\pm1$ \\
HD2885\tnote{d} & 00:31:33.47 & -62:57:56.02 & A2V & 8600 & $53\pm10$ \\
HD30051  &04:43:17.20 & -23:37:42.06 & F2/3IV/V & 6600 & $58\pm4$ \\
HD53842   &06:46:13.54 & -83:59:29.51 & F5V & 6600 & $57\pm2$ \\
HD1466  &00:18:26.12 & -63:28:38.98 & F9V & 6200
 & $41\pm1$\\
HD105 & 00:05:52.54 & -41:45:11.04 & G0V & 6000 & $40\pm1$ \\
HD12039 & 01:57:48.98 & -21:54:05.35 & G3/5V & 5600 & $42\pm2$ \\
HD202917 & 21:20:49.96 & -53:02:03.14 & G5V & 5400 & $46\pm2$ \\
HD44627\tnote{d} & 06:19:12.91 & -58:03:15.52 & K2V & 5200 & $46\pm2$ \\
HD55279 & 07:00:30.49 & -79:41:45.98 & K3V & 4800 & $64\pm4$ \\
HD3221 & 00:34:51.20 & -61:54:58.14 & K5V & 4400 & $46\pm2$ \\
HIP107345 & 21:44:30.12 & -60:58:38.88 & M1 & 3700 & $42\pm5$ \\
HIP3556 &00:45:28.15 & -51:37:33.93 & M1.5 & 3500 & $39\pm4$ \\
GSC8056-482 &02:36:51.71 & -52:03:03.70 & M3Ve & 3400 & 25\\
\hline
\end{tabular}
\begin{tablenotes}
\item[a]{Spectral types listed are from the SIMBAD Astronomical Database}
\item[b]{Calculated from stellar modeling.  See Section 3}
\item[c]{Distances are taken from the {\it Hipparcos} Catalog \citep{Perryman97}}
\item[d]{Binary or multiple star system}
\end{tablenotes}
\end{threeparttable}

\end{table*}

The 30 Myr-old Tucana-Horologium Association is the oldest in the GASPS survey. The Tucana-Horologium Association, discovered independently by \citet{Zuckerman00} and \citet{Torres00}, is a group of $\sim60$ stars with common proper motion and an average distance of 46 pc \citep{Zuckerman04}.  About $\sim1/3$ of the targets have debris disk systems known from previous {\it Spitzer} surveys \citep{Hillenbrand08, Smith06}.

We obtained {\it Herschel} PACS photometry of the seventeen GASPS targets in the Tucana-Horologium Association.  We also obtained PACS spectra for two of the targets.  Previously unpublished {\it Spitzer} IRS spectra for three targets are presented.  In Section 2, we present our methods and results of data reduction and aperture photometry.  In Section 3, we fit blackbody and modified blackbody models to the detections and upper limits to determine dust temperatures and fractional luminosities.  We further analyze some of the disks with our optically thin dust disk model in Section 4.  Additionally, we discuss the detection of a marginally resolved disk in our sample in Section 5 and present conclusions in Section 6.

\section{Observations and Data Reduction \label{obs}}

The stellar properties of the seventeen Tucana-Horologium stars observed are listed in Table \ref{tab:stars}.  These stars have spectral types ranging from B9-M3 and distances of 25-64 pc.  
PACS scan map observations were obtained for all seventeen Tucana-Horologium targets at 70 and 160$\,\mu$m.  Additional observations at 100$\,\mu$m were taken for five targets.  The stars were observed at two scan angles, $70\deg$ and $110\deg$. The two scans were combined to reduce the excess noise caused by streaking in the scan direction, as suggested by the Herschel PACS Instrument Calibration Centre (ICC)\footnotemark[12].  Each scan map was executed with the medium scan speed (20$^{\prime\prime}$s$^{-1}$) and consisted of 10 legs with scan lengths of 3$^{\prime}$ and scan leg separation of $4^{\prime\prime}$.  The number of repetitions varied from 1 to 4 based on the expected flux density of the star.  The on-source time for each observation is given in Table \ref{tab:hipe7}.  Two targets, HD105 and HD3003, were also observed in the PACS LineScan mode at 63$\,\mu$m and 190$\,\mu$m, targeting the OI fine structure line and DCO+ respectively.

\footnotetext[12]{PICC-ME-TN-036: \url{http://herschel.esac.esa.int/twiki/pub/Public/\newline PacsCalibrationWeb/PhotMiniScan\_ReleaseNote\_20101112.pdf}}

\begin{table*}[!ht]
\centering
\caption{Herschel PACS Photometry Results\label{tab:hipe7}}
\begin{threeparttable}
\begin{tabular}{|l| c c c | c c c | c c c |}
\hline\hline
 & \multicolumn{3}{|c|}{70$\,\mu$m} & \multicolumn{3}{|c|}{100$\,\mu$m}& \multicolumn{3}{|c|}{160$\,\mu$m} \\
\hline
Star  & Obs. Flux\tnote{$\ast$} & Stellar Flux & On-source & Obs. Flux  & Stellar Flux &On-source& Obs. Flux  & Stellar Flux &On-source\\
& (mJy) & (mJy) &Time (s)& (mJy) & (mJy)&Time (s)& (mJy) & (mJy)&Time (s) \\
\hline
\multicolumn{10}{|c|}{Debris Disks}\\
\hline
HD105 & $128.3\pm 7.0$ & 3.1 & 72	& $151.2\pm5.7$ & 1.7& 144	&$81.2\pm 12.3$ & 0.6 & 144 \\
HD3003 & $59.7\pm3.8$ & 7.6 & 72	& $19.0\pm2.4$ & 4.1 & 144	&$<18.2$ & 1.5 & 144\\
HD1466 & $13.0\pm0.9$ & 2.8 & 720	& -- & -- & --			&$<10.6$ & 0.6 & 720\\
HD30051 & $23.4\pm 1.1$ & 3.2 & 720	& -- & -- &--			&$16.8\pm 2.4$ & 0.6 & 720\\
HD202917 & $33.9\pm1.6$ & 1.6 & 360	& $29.9\pm2.5$ & 0.8 & 144	&$17.7\pm 3.8$ & 0.3 & 360\\
HD12039 & $10.5\pm0.8$ & 2.3 & 720	& -- & --& --			&$<15.5$ & 0.5 & 720\\
\hline
\multicolumn{10}{|c|}{Non-Excess Stars}\\
\hline
HD2884 & $7.7\pm1.4$ & 12.1 & 360	& -- & --& --			&$<18.6$ & 2.3 & 360\\
HD16978 & $15.4\pm 1.7$ & 15.8 & 144	& -- & --& --			&$<14.4$ & 3.1 & 144\\
HD224392 & $8.6 \pm 1.1$ & 8.7 & 540	& -- & -- & --			&$<10.0$ & 1.7 & 540\\
HD2885 & $16.0\pm 1.4$ & 16.5 & 360	& -- & -- & --			&$<8.5$ & 3.3  & 360\\
HD53842 & $9.9\pm3.3$ & 2.3 & 72	& $<9.2$ &1.3& 144		&$<15.4$ & 0.5 & 144\\
HD44627 & $<3.5$& 1.4 & 360		& -- & --& --			&$<10.3$ & 0.3 & 360\\
HD55279 & $<3.1$ & 0.9 & 360		& -- & --& --			&$<8.5$ & 0.1  & 360\\
HD3221 & $<3.7$ & 2.4 & 360		& -- & --& --			&$<10.4$ & 0.4 & 360\\
HIP107345 & $<3.3$ & 0.9 & 360		& -- & -- & --			&$<10.8$ & 0.2 & 360\\
HIP3556 & $<16.5$ & 1.3 & 72		& $<16.9$ & 0.7 & 144		&$<17.4$& 0.2 & 144\\
GSC8056-482 & $<3.6$ & 1.6 & 360	& -- & -- & --			&$<20.4$ & 0.3 & 360\\
\hline
\end{tabular}
\begin{tablenotes}
\item[$\ast$]{1$\sigma$ uncertainties include statistical and systematic errors added in quadrature.  The expected stellar fluxes at the PACS wavelengths were calculated by fitting the optical and near-IR data with NextGen stellar atmosphere models \citep{NextGen}.}
\end{tablenotes}
\end{threeparttable}
\end{table*}

\subsection{Photometry}
The data were reduced with HIPE 7.2 \citep{Ott10} using the standard reduction pipeline. The final maps have pixel scales of 3.2/3.2/6.4$^{\prime\prime}$pixel$^{-1}$ in the 70/100/160$\,\mu$m images respectively, corresponding to the native pixel scales of the PACS detectors. The two scans were reduced separately, then averaged together.  The flux values were measured using an IDL aperture photometry code with apertures of 5.5/5.6/10.5$^{\prime\prime}$ for the 70/100/160$\,\mu$m images respectively, as recommended for faint sources. Aperture corrections were applied based on the encircled energy fraction from PACS observations of Vesta provided by the Herschel PACS ICC\footnotemark[13].  The sky annulus for error estimation was placed 20-30$^{\prime\prime}$ away from star center for the 70 and 100$\,\mu$m images and 30-40$^{\prime\prime}$ away for the 160$\,\mu$m images. For three of the targets, HD2884, HD2885 and HD53842, the fields were contaminated by nearby stars and background galaxies, so the sky annulus was offset to a nearby clean field.  

The RMS pixel uncertainty, $\sigma_{pix}$, was estimated by calculating the standard deviation of the pixels in the sky annulus. The total statistical error in the measurement is given by
\begin{equation} \sigma_{tot} = \frac{\sigma_{pix}}{\alpha_{corr} x_{corr}} \sqrt{n_{ap}\left(1+\frac{n_{ap}}{n_{sky}}  \right)}
,\end{equation}
where $\alpha_{corr}$ is the aperture correction, $x_{corr}$ is the correlated noise correction factor\footnotemark[13] (0.95 at 70 and 100$\,\mu$m, 0.88 at 160$\,\mu$m), and $n_{ap}$ and $n_{sky}$ are the number of pixels in the aperture and annulus respectively.  
An absolute calibration error was also added in quadrature with the statistical error to give the total uncertainty reported in Table \ref{tab:hipe7}. The absolute calibration errors given by the Herschel PACS ICC are 2.64/2.75/4.15\% for the 70/100/160$\,\mu$m images respectively\footnotemark[13].  Upper limits for all non-detections were determined using the 3$\sigma$ errors from the same aperture photometry method as above.

\footnotetext[13]{PICC-ME-TN-037: \url{http://herschel.esac.esa.int/twiki/pub/Public/\newline PacsCalibrationWeb/pacs\_bolo\_fluxcal\_report\_v1.pdf}}

\begin{figure}
\epsscale{1.2}
\plotone{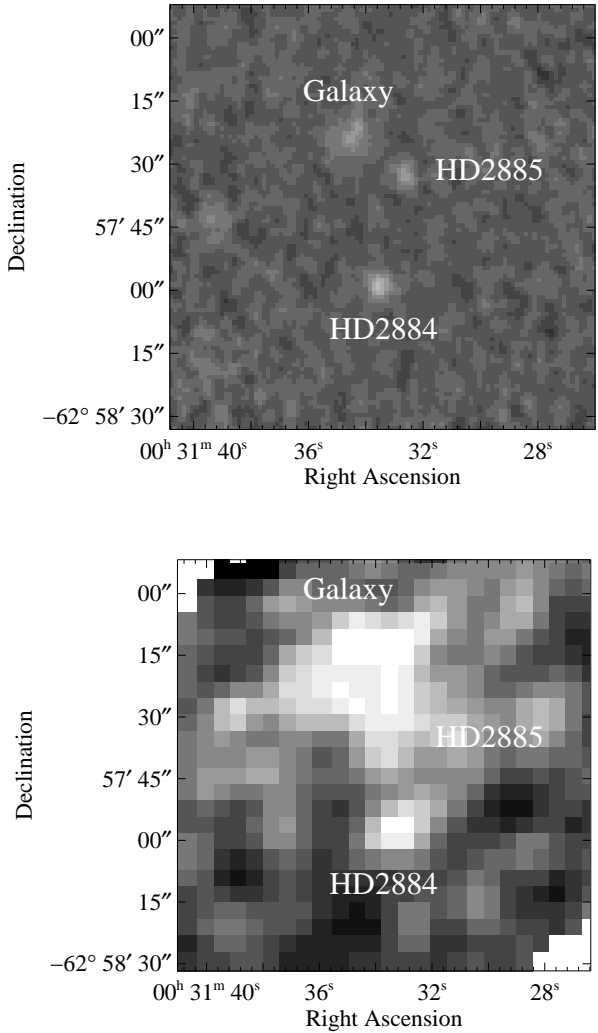}
\caption{The field of HD2884 and HD2885 at 70$\,\mu$m with {\it Herschel} (top) and {\it Spitzer} (bottom).  The better spatial resolution of {\it Herschel} allows the sources to be cleanly separated and avoids contamination of photometry. \label{fig:contam}}
\end{figure}

The PACS photometry fluxes and upper limits are listed in Table \ref{tab:hipe7}.  Six of the seventeen targets were determined to have infrared excesses above the photosphere, including one previously unknown disk, HD30051.  A significant excess is defined here as a 70$\,\mu$m photosphere-subtracted flux greater than 3$\sigma$.  One of our targets, HD2884, was previously suspected to have an infrared excess \citep{Smith06}.  In the PACS data, this target seems to have less flux than the expected photospheric value at 70$\,\mu$m.  The {\it Spitzer} IRS spectrum also shows no hint of an IR excess. \citet{Smith06} notes the possibility of contamination of HD2884 with a background galaxy, but could not separate the two sources with {\it Spitzer}'s spatial resolution.  Figure \ref{fig:contam} shows the confused field of HD2884 observed with {\it Herschel} and {\it Spitzer} at 70$\,\mu$m.  {\it Herschel}'s improved spatial resolution allows us to properly resolve the sources and avoid contamination. We determine that HD2884 has no detectable debris disk.

\begin{figure}
\epsscale{1.2}
\plotone{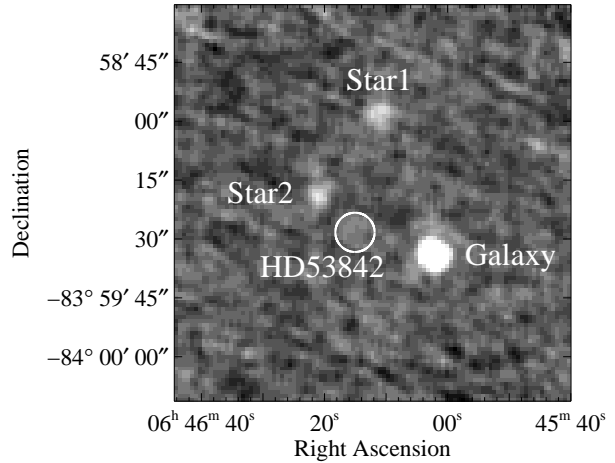}
\caption{The field of HD53842 at 70$\,\mu$m with {\it Herschel}.  The star is barely detected at 70$\,\mu$m ($3\sigma$ detection). The nearby sources may have contaminated the {\it Spitzer} IRS spectrum of HD53842.\label{fig:HD53842}}
\end{figure}

HD53842 has a photosphere subtracted flux that is only 2.3$\sigma$ above the noise at 70$\,\mu$m.  The {\it Spitzer} IRS spectrum from \citet{Moor09} also shows a marginal excess at 14-35$\,\mu$m but may suffer from contamination.  The {\it Spitzer} 24$\,\mu$m image shows two nearby stars within 15$^{\prime\prime}$ whose Point Spread Functions (PSFs) appear to overlap with that of HD53842. These sources are also seen in the PACS images (see Fig.\ \ref{fig:HD53842}) and make an accurate determination of the flux difficult. We are unable to confirm if HD53842 has a debris disk.  

\subsection{Spectroscopy}

Two targets, HD105 and HD3003, were observed in the LineScan mode of the PACS instrument.  The lines targeted were [OI] at 63.185$\,\mu$m and DCO+ at 189.570$\,\mu$m.  The continuum detection limit was 202 mJy at 63$\,\mu$m.  The photometry detection at 70$\,\mu$m was below this limit for both HD105 and HD3003, and the noise levels in the spectra are comparable to or larger than the flux at $70\,\mu$m.  Therefore, we conclude that the continuum is not detected in these observations.  There are also no emission lines detected above the noise.

Upper limits to the line fluxes were calculated by integrating over a Gaussian with a width equal to the instrumental Full-Width Half-Maximum (FWHM) and a height given by the standard deviation of the noise, placed at the wavelength where the line is expected.  The upper limits are given in Table \ref{tab:lines}.

The three targets with previously unpublished {\it Spitzer} spectra (HD1466, HD2884 and HD3003) were observed with the IRS spectrograph \citep{Houck04}, using the Short-Low (5.2-14$\,\mu$m) and Long-Low modules (14-38$\,\mu$m; $\lambda/\Delta\lambda \sim 60$).  These three systems were observed as part of the {\it Spitzer} GTO program 40651 (PI: J. Houck).  Since the sources were expected to be unresolved and the pointing of {\it Spitzer} was excellent, all of our targets were observed in Staring mode with no peak-up.  We carried out the bulk of the reduction and analysis of our spectra with the IRS team's SMART program \citep{Higdon04}.

\begin{table}
\begin{center}
\caption{Herschel Spectroscopy: $3\sigma$ Line Upper Limits\label{tab:lines}}
\begin{tabular}{l c c c}
\tableline\tableline
Star  & Line & Wavelength & $3\sigma$ Upper Limit\\
      &		& ($\,\mu$m)	& (W/m$^2$) \\
\tableline
HD105 & [OI] & 63.185	& $<9.63\times10^{-18}$ \\
      & DCO+ & 189.570 & $<1.18\times10^{-17}$ \\
HD3003& [OI] & 63.185	& $<1.43\times10^{-17}$ \\
      & DCO+ & 189.570 & $<8.55\times10^{-18}$ \\
\tableline
\end{tabular}
\end{center}
\end{table}

\section{Blackbody and Modified Blackbody Fits\label{bbfits}}
\begin{figure*}
\epsscale{1.13}
\plotone{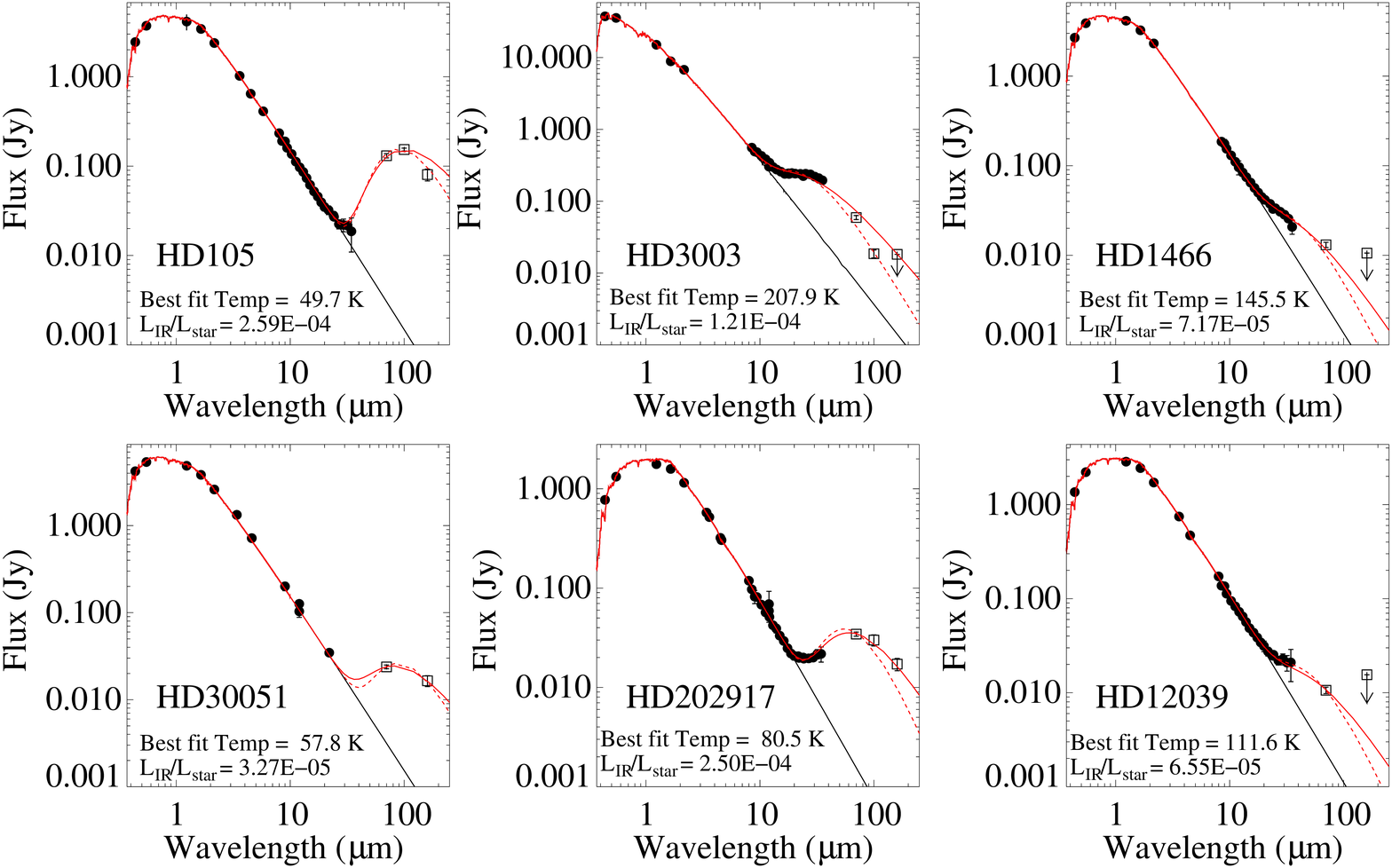}
\caption{Single temperature blackbody fits to the SEDs of the 6 debris disks in the sample.  Filled circles are data from the literature, open squares are the {\it Herschel} PACS data.  The solid black curve is the fit to the stellar photosphere using the NextGen models \citep{NextGen}.  The red curve shows the best fitting single temperature blackbody model.  The figure label for each disk displays the best fitting temperature and the fractional IR luminosity L$_{\text{IR}}$/L$_\ast$. The dashed red lines show the best-fitting modified blackbody models.  Available MIPS 70$\,\mu$m data are not plotted here, as they are consistent with the PACS data, which have smaller uncertainties.  \label{fig:seven}}
\end{figure*}

\begin{figure*}
\plotone{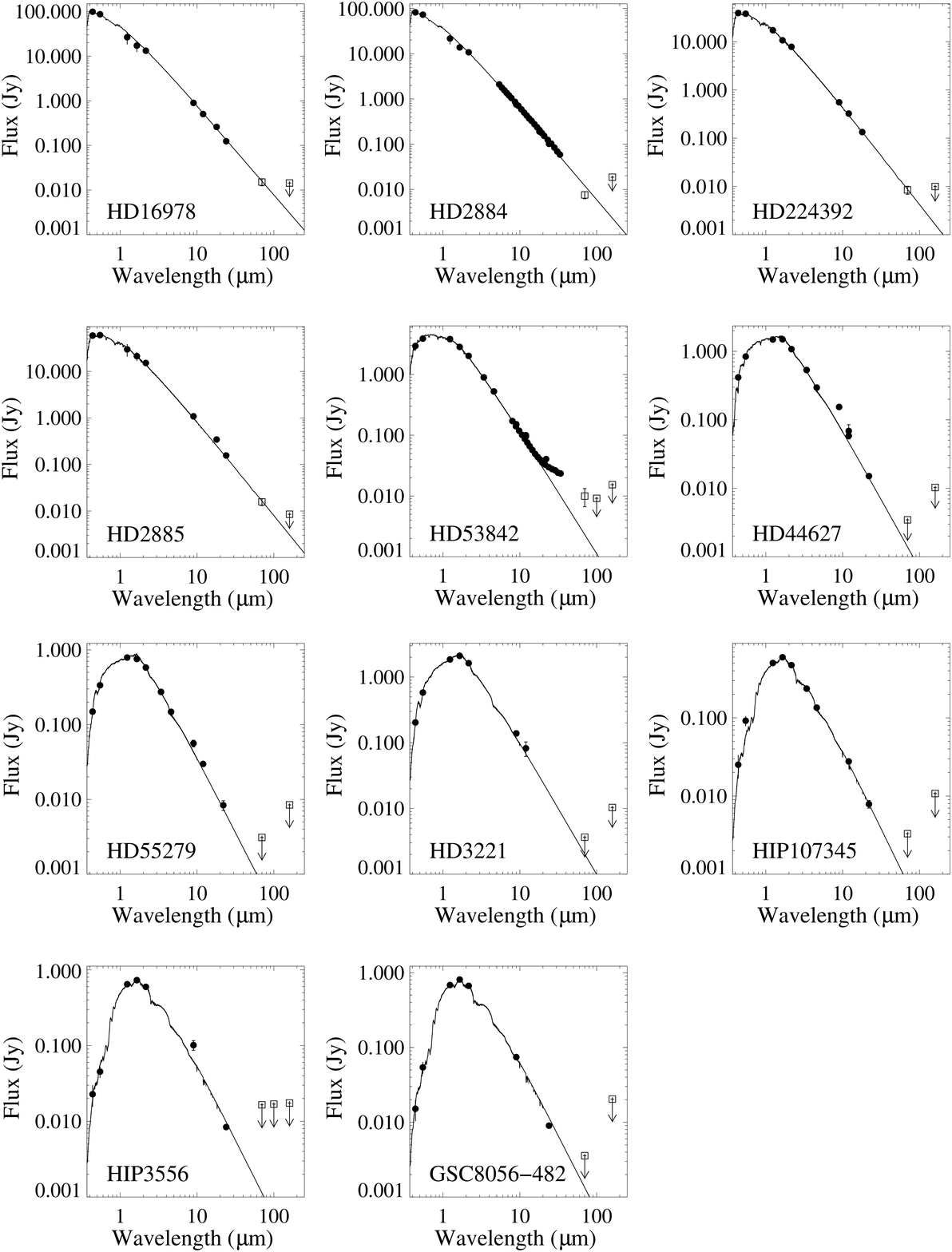}
\caption{Spectral energy distributions for the 11 stars that were not found to have debris disks.  The solid curve is the fit to the stellar photosphere using the NextGen models \citep{NextGen}. Upper limits for L$_{\text{IR}}$/L$_\ast$ are shown in Table \ref{tab:bbfit}. Available MIPS 70$\,\mu$m data are not plotted here, as they are consistent with PACS data for every source except HD2884, which lies in a confused field (see Fig.\ \ref{fig:contam}).  HD53842 has only a marginal detection of an excess at 70$\,\mu$m and also lies in a confused field (see Fig.\ \ref{fig:HD53842}).  Consequently, we are unable to determine if HD53842 has a debris disk.\label{fig:bb1}}
\end{figure*}

We constructed spectral energy distributions for all sources using data from {\it Hipparcos, 2MASS, Spitzer} (IRAC, IRS, and MIPS), {\it AKARI, IRAS}, the {\it WISE} preliminary release, and our new {\it Herschel} data.  The SEDs of debris disk stars are shown in Figure \ref{fig:seven} and those of non-excess stars in Figure \ref{fig:bb1}.  The data used for these figures are listed in Table \ref{tab:archive} (online version).  For each target, we fitted the stellar photosphere with NextGen models of stellar atmospheres \citep{NextGen}. The best fitting stellar model was determined through $\chi^2$ minimization with effective temperature and the normalization factor as free parameters. The {\it Spitzer} IRS spectra were binned to a resolution of $\Delta\lambda/\lambda \sim 0.1$ and only the data with $\lambda > 8\,\mu$m were used to determine the fit of the excess. 

For our six disk detections, we fitted the excess emission with a standard single temperature blackbody model with two parameters, the temperature and the fractional disk luminosity.  This model is a simplified disk representation that assumes all the dust is at the same temperature, and behaves like a perfect blackbody.  While this model is simple, it gives a good first estimate of the dust temperature and abundance, and unlike more physical models, one can get a fit even with very few data points.  The best fit was determined through $\chi^2$ minimization.  We also calculated the goodness-of-fit for each best fitting model for comparison with more detailed models described in Section 4.  The goodness-of-fit, $Q$, is the probability that a $\chi^2$ value this poor will occur by chance given the error in the data.  
$Q$ is given by the normalized incomplete gamma function \begin{equation} Q(a,x) = \frac{\Gamma(a,x)}{\Gamma(a)}\; ; \quad a = \frac{N}{2} \quad x = \frac{\chi^2}{2}, \end{equation} where $\Gamma(a,x)$ and $\Gamma(a)$ are incomplete and complete gamma functions respectively, and $N$ is the number of degrees of freedom.
$Q$ varies from 0 to 1, with larger values indicating better fits. An acceptable model is one that has a value of $Q\gtrsim10^{-3}$ \citep{Numerical}.

The $Q$ values, however, are largely influenced by the IRS spectrum.  The IRS spectra have more data points compared to the few PACS photometry points.  Therefore, fits that may appear to be just as good, can have very different $Q$ values if the uncertainties in the IRS spectra are different.  The point of the $Q$ value is not to compare the goodness-of-fit between disks, but to compare between models of a given disk.  

\begin{table}[!h]
\centering
\caption{Model Fit Results\label{tab:bbfit}}
\begin{tabular}{l c c c}
\hline\hline
\multicolumn{4}{c}{Debris Disks}\\
\hline
\multicolumn{4}{c}{Blackbody Fits}\\
Star  & Temp (K) & L$_{\text{IR}}$/L$_\ast$ & Q\\
\hline
HD105    &  $49.7\pm1.4$	& $(2.59\pm0.05)\times10^{-4}$  & 0.004	\\
HD3003	 &  $207.9\pm2.5$	& $(1.21\pm0.01)\times10^{-4}$  &   0.00	\\
HD1466   &  $145.5\pm6.7$	& $(7.21\pm0.11)\times10^{-5}$  &   0.61\\
HD30051  &  $57.8\pm6.4$	& $(3.27\pm0.20)\times10^{-5}$	&   0.82  \\
HD202917 &  $80.5\pm1.5$	& $(2.50\pm0.01)\times10^{-4}$	&   0.93  \\
HD12039  &  $111.6\pm6.2$	& $(6.55\pm0.39)\times10^{-5}$  &   0.38 \\
\hline
\multicolumn{4}{c}{Modified Blackbody Fits}\\
\hline
HD105    &  $40.3\pm1.2$	& $(2.41\pm0.90)\times10^{-4}$  & 0.11	\\
HD3003	 &  $160.6\pm1.6$	& $(1.08\pm0.01)\times10^{-4}$  &   0.00\\
HD1466   &  $106.8\pm4.3$	& $(6.33\pm0.03)\times10^{-5}$  &   0.002	\\
HD30051  &  $39.8\pm2.8$	& $(2.81\pm0.05)\times10^{-5}$	&   0.82  \\
HD202917 &  $64.6\pm1.1$	& $(2.50\pm0.01)\times10^{-4}$	& 0.13  \\
HD12039  &  $85.1\pm3.7$	& $(6.33\pm0.33)\times10^{-5}$  &   0.43 \\
\hline
\multicolumn{4}{c}{Non-Excess Stars}\\
\hline
	  & \multicolumn{1}{c}{Blackbody} & Modified Blackbody 	& \\
	  & \multicolumn{1}{c}{L$_{\text{IR}}$/L$_\ast$} & L$_{\text{IR}}$/L$_\ast$ & \\
\hline
HD2884	 &  $<2.9\times10^{-6}$ & $<6.7\times10^{-6}$		& \\
HD16978	 &  $<1.9\times10^{-6}$ & $<8.6\times10^{-7}$		& \\
HD224392 &  $<5.0\times10^{-6}$ & $<1.6\times10^{-5}$		& \\
HD2885	 &  $<2.7\times10^{-6}$ & $<1.8\times10^{-5}$		& \\
HD53842  &  $<6.4\times10^{-5}$ & $<2.9\times10^{-4}$		& \\
HD44627	 &  $<1.9\times10^{-4}$ & $<1.7\times10^{-4}$		& \\
HD55279	 &  $<8.6\times10^{-4}$ & $<2.5\times10^{-3}$		& \\
HD3221	 &  $<2.4\times10^{-4}$ & $<1.3\times10^{-3}$		& \\
HIP107345&  $<8.4\times10^{-4}$ & $<4.2\times10^{-4}$		& \\
HIP3556	 &  $<1.2\times10^{-3}$ & $<5.4\times10^{-4}$		& \\
GSC8056-482&$<1.2\times10^{-3}$ & $<5.3\times10^{-4}$		& \\
\hline
\end{tabular}
\end{table}

From the fits, we determined the disk temperatures and the fractional infrared luminosities (L$_{\text{IR}}$/L$_\ast$).  For the remaining sources, we determined upper limits on L$_{\text{IR}}$/L$_\ast$ by fitting blackbody models to the flux upper limits.  Upper limits on L$_{\text{IR}}$/L$_\ast$ depend on temperature, so the largest L$_{\text{IR}}$/L$_\ast$ value found assuming dust temperatures from 10-300 K was adopted.  
The results of the fits to the disks are given in Table \ref{tab:bbfit}.  The disks display a large range of temperatures and L$_{\text{IR}}$/L$_\ast$ values, showing no correlation with spectral type. 

We also fit the disks with a modified blackbody model.  This model assumes that at longer wavelengths, the grains no longer emit as blackbodies, but instead the dust emissivity is decreased by a factor $(\lambda/\lambda_0)^{-\beta}$ for $\lambda > \lambda_0$, due to a lower dust opacity at longer wavelengths compared to a blackbody.  Here $\lambda_0 = 2\pi a_{blow}$, where $a_{blow}$ is the blowout size due to radiation pressure.  We do not have sub-mm or mm data to constrain the parameter $\beta$, therefore we assume $\beta = 1.0$ for these models based on previous measurements for debris disks \citep{Dent00,Williams06,Nilsson10}.  ISM dust has $\beta\approx2.0$ \citep{Boulanger96}, but debris disks are expected to have lower values of $\beta$ due to the presence of larger grains \citep{Draine06}.  The results of the modified blackbody fits are also shown in Table \ref{tab:bbfit}.  The use of modified blackbody models shifts the characteristic temperatures to lower values, a trend which was also noticed by \citet{Carpenter09}.  The goodness-of-fit was improved for HD105, HD3003, and HD12039 with the use of the modified blackbody model, but made worse for HD1466 and HD202917.

The blackbody and modified blackbody models provide a first look at the disk properties.  For example, disks with low characteristic temperature such as HD105, HD30051, and HD202917, are likely to have large inner gaps and copious amounts of cold dust far from the star.  To test this idea, we need a more physical model of the disk.

\section{Dust Disk Model\label{model}}
To further investigate the disk properties, we fit the disks with an optically thin dust disk model.  Rather than assuming the grains to be perfect blackbodies, the dust model assumes the grains have a particular emissivity that is dependent on the size of the dust grains and the wavelength of radiation being absorbed or emitted.  For the Tucana-Horologium disks, we make the assumption that the grains are purely silicate in composition, specifically, astronomical silicates \citep{Draine84, Laor93, Weingartner01}.  The poorly populated SEDs and lack of resolved imaging prevents us from fitting more complex grain compositions.  

The geometry of the disks is described by three parameters: the inner and outer radius ($r_{min}$ and $r_{max}$) and the radial surface density profile.  The azimuthally symmetric radial surface density profile is characterized by a power-law with index $q$, such that the surface density $\Sigma$ varies with radius as $\Sigma(r)dr \propto r^q dr$.  A power-law index of $q=0$ is expected for a transport dominated disk, and an index of $q=-1.5$ is expected for a collisionally dominated disk \citep{Krivov06,Strubbe06}.

The population of dust grains is also characterized by three parameters: the minimum and maximum dust grain sizes in the disk ($a_{min}$ and $a_{max}$), and the distribution of grains with sizes between $a_{min}$ and $a_{max}$.  The grain size distribution is typically assumed to be a power-law with index $\kappa$ such that the number density varies with grain size $a$ as $n(a)da \propto a^\kappa da$, where a value of $\kappa \sim -3.5$ is expected for a steady-state collisional cascade \citep{Dohnanyi69}.

The model iteratively determines the equilibrium dust temperature at each radius for each grain size by balancing radiation absorbed from the star with radiation re-emitted through the formula
\begin{align} \int_0^\infty Q_{abs}(\nu, a)& \left(\frac{R_\ast}{r}\right)^2 B_\nu(T_{eff}) d\nu = \nonumber \\ &\int_0^\infty 4 Q_{abs}(\nu,a)B_\nu(T_d(r,a))d\nu, \end{align} 
where $B_\nu(T_{eff})$ is the blackbody flux coming from the surface of the star with temperature $T_{eff}$ and radius $R_\ast$, $B_\nu(T_d(r,a))$ is the blackbody flux radiating from the dust grain of size $a$ at a distance $r$ away from the star, and $Q_{abs}(\nu,a)$ is the dust absorption coefficient calculated by \citet{Draine84} for an astronomical silicate grain of size $a$.  The total flux is then determined by summing up over all radii and grain sizes according to the formula

\begin{align}
  F_\nu& = \,A \int_{a_{min}}^{a_{max}} \left(\frac{a}{a_{max}}\right)^{\kappa} da \nonumber \\
	   &\cdot \int_{r_{min}}^{r_{max}} 4 \pi a^2 Q_{abs}(\nu,a) \pi B_{\nu}(T_d(r,a))  2 \pi r \left(\frac{r}{r_{min}}\right)^q dr,
\end{align}
where $A$ is a normalization constant that includes the distance to the system and the total amount of material in the disk.

\subsection{Model Parameters}

The Tucana-Horologium debris disks are faint (all are under 150 mJy at 70$\,\mu$m), so the SEDs of these targets are not well sampled due to lack of detections at the longer wavelengths.  For this reason, we must fix some of the model parameters.  In particular, because of the lack of sub-mm and longer wavelength data, we are unable to constrain parameters that affect this region of the SED, specifically cold grains that are larger than 1 mm or farther out than $\sim 120$AU. Therefore, we fix the maximum grain size and the outer radius to these values. For HD202917, we fixed the outer radius to 80 AU to be consistent with HST imaging \citep{Krist07,Mustill09}. The radial surface density profile also cannot be constrained with the current data.   
Hence we fix the power-law index of the radial density profile to $q=-1.5$, the value expected for collisionally dominated disks.  

The SEDs of HD105, HD3003, and HD202917 are populated enough to get well constrained fits with only these three parameters fixed.  However, HD1466 and HD12039 were only detected at $70\,\mu$m with PACS.  The models for these disks were further constrained by fixing the minimum grain size and the grain size distribution.  The grain size distribution was fixed to a power-law index of $\kappa = -3.5$, the value expected for a steady-state collisional cascade, and the minimum grain size was fixed to the expected blowout size for astronomical silicates.  This is the size at which the radiation pressure is half of the gravity; grains this size and smaller are ejected from the system.  The blowout size is calculated from the formula given in \citet{Backman93} as \begin{equation} \left(\frac{a_{blow}}{1\,\mu m}\right) = 1.15 \left(\frac{L_\ast}{L_\odot}\right) \left(\frac{M_\odot}{M_\ast}\right) \left(\frac{1\text{g cm}^{-3}}{\rho}\right) \end{equation} assuming constant density, spherical grains produced from planetesimals on circular orbits.  This equation assumes that the grains are of a certain composition and all have the same density, and that the density is uniform throughout the grain. For astronomical silicates, we assume a grain density of 2.5 g cm$^{-3}$.  This simple equation provides a good estimate of the expected minimum grain size, but it relies heavily upon many assumptions and must be treated solely as an estimate, accurate perhaps to only an order of magnitude.

\subsection{Results}

\begin{figure*}
\epsscale{1.2}
\plotone{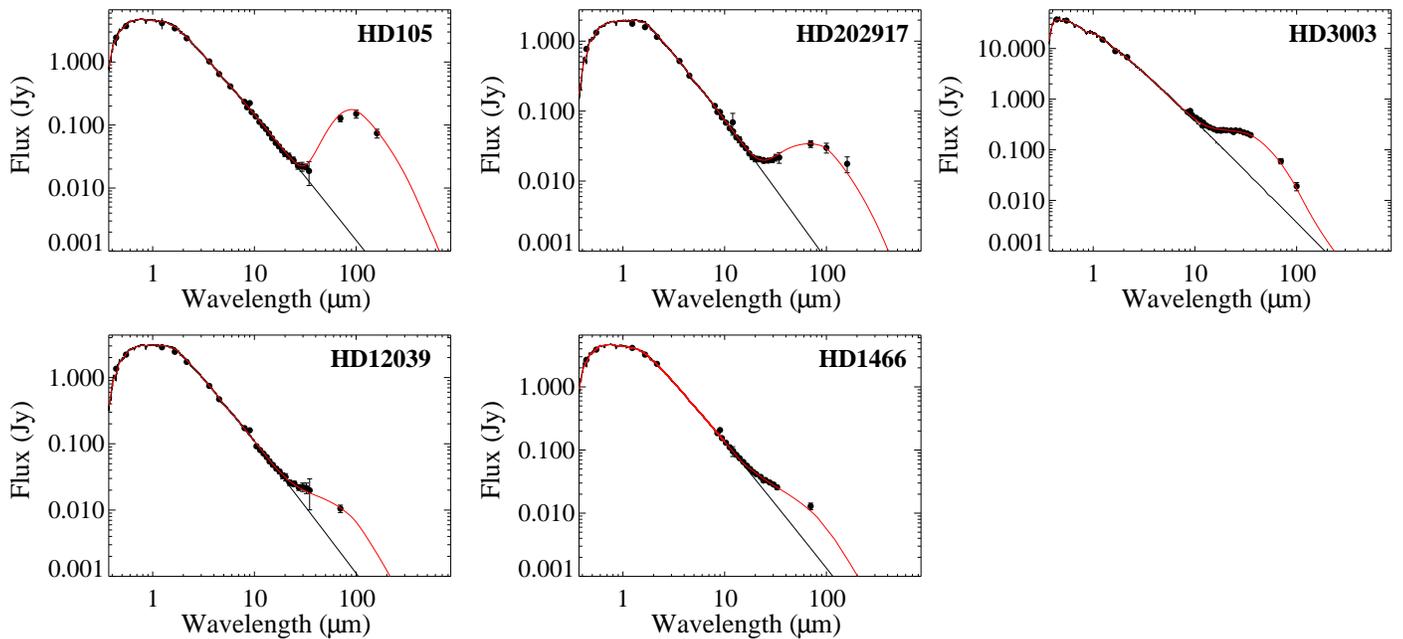}
\caption{Best fitting models for the five of the debris disks with well populated SEDs.  The solid black lines shows the stellar photospheres that are fit with the NextGen models \citep{NextGen}.  The best fitting models of the disks are shown in red.  The best fitting model parameters are given in Table \ref{tab:modparms}. \label{fig:models}}
\end{figure*}

To determine the best model for each system, we compared a grid of models to the SED and found the set of parameters that produced the $\chi^2$ minimum value. The error of each fit parameter was determined from the $1\sigma$ confidence interval in the $\chi^2$ distribution after fixing the other parameters to their best fit values. 
The best fitting parameters and the errors are displayed in Table \ref{tab:modparms}, and the best fitting SED models are shown in Figure \ref{fig:models}.  The HD30051 disk could not be fit at all with this model because its SED has too few data points, having not been known to have a debris disk before {\it Herschel}.  

Lower limits on the disk masses are also given in Table \ref{tab:modparms}.  These masses are calculated by summing up the mass of silicate grains with an assumed uniform density of 2.5 g cm$^{-3}$, with sizes between $a_{\text{min}}$ and $a_{\text{max}}$ and disk radii between $r_{\text{min}}$ and $r_{\text{max}}$.  The mass also depends on the normalization constant and the radial density profile of the disk.  It is a lower limit on the mass because it only takes into account the dust smaller than 1 mm, and not larger pebbles and planetesimals hidden in the disk.  

\subsubsection{HD105}
HD105, a G0V star 40 pc away, has the largest IR excess in the PACS wavebands of all the Tucana-Horologium disks in this sample, yet shows very little mid-IR excess in the {\it Spitzer} IRS spectrum.  This disk was observed with {\it Spitzer} as part of the Formation and Evolution of Planetary Systems (FEPS) Legacy Survey \citep{Meyer06}.  \citet{Meyer04} first fit the disk with models from \citet{Wolf03} that used \citet{Draine84} grain emissivities for astronomical silicates and graphite in ISM ratios and found an inner radius of 45 AU for a fixed minimum grain size of 5$\,\mu$m.  They assumed a flat radial density profile (q=0), a grain size distribution power-law index of $\kappa=-3.5$, an outer radius of 300 AU, and a maximum grain size of 100$\,\mu$m.  

\citet{Hollenbach05} found an even lower $\chi^2$ value using the \citet{Wolf03} models with a minimum grain size of 21$\,\mu$m and a lower inner radius of 19 AU, but confirmed that the inner cavity must be $\geq13$ AU because of the lack of IR excess at $\lambda \lesssim35\,\mu$m.  The {\it Spitzer} IRS spectrum also shows no evidence of gas lines, and \citet{Hollenbach05} determined the gas mass to be $<0.1\,\text{M}_{\text{J}}$ between 10--40 AU for a constant surface density.

Later FEPS modeling by \citet{Hillenbrand08} used multi-temperature blackbodies to fit the HD105 disk, and found an inner radius of 36.8 AU.  This is a lower limit to the inner radius since the grains were assumed to be large blackbody grains. Smaller grains could reach the same temperatures farther from the star.

The smaller error bars of the {\it Herschel} PACS measurements, and the addition of the 100$\,\mu$m data point, better map the characteristic turnover point in the SED and lessens the severity of the model degeneracy.  
Our results are consistent with those of \citet{Meyer04}, \citet{Hollenbach05}, and \citet{Hillenbrand08}.

The lack of mid-IR excess in this disk indicates an absence of both small grains and grains within a large inner cavity.  The dust begins at about the distance of the Solar System's Kuiper belt (52 AU using our best fitting model) and extends to some distance beyond (see Section 6 for an estimate of the outer radius of HD105).  The lack of detectable dust interior to this region may be due to one or more planets orbiting inside this dust ring \citep{Moro-Martin&Molhotra05}. \citet{Apai08} searched for massive planets in the HD105 disk using VLT/NACO angular differential imaging and found none. However, the survey was only sensitive to planets with masses $>6$ M$_J$ at distances $>15$ AU.  

The minimum grain size for this disk (8.9$\,\mu$m) is much larger than the expected blowout size of 0.5$\,\mu$m.  This suggests either the calculation of the blowout size is inaccurate, or the small grains are efficiently removed by some other mechanism.  The calculation of the blowout size is just an estimate with several assumptions built in, such as grain density and composition, and a change in the assumptions could lead to a drastically different blowout size.  For instance, if the grains were porous, they would be less massive at a certain size, and the larger grains could be more easily blown out of the system.

\begin{table*}[!ht]
\centering
\caption{Best Fitting Model Parameters\label{tab:modparms}}
\begin{threeparttable}
\begin{tabular}{l c c c c c c c c c}
\hline\hline
Target	& $a_{min}$\tnote{a}	& $a_{blow}$	& $a_{max}$\tnote{a}	&$\kappa$\tnote{a}	& $r_{min}$\tnote{a}	& $r_{max}$\tnote{a}	&$q$\tnote{a}	&Q\tnote{b}	& Mass (M$_\oplus$)\tnote{c} \\
\hline
HD105	& $8.9^{+11.1}_{-4.4}\,\mu$m &0.5$\,\mu$m	& {\it 1000$\,\mu$m}	&$-3.3^{+0.3}_{-0.4}$	& $52^{+36}_{-7}$AU	& {\it 120 AU}		& {\it-1.5}	& 0.15	& $4.3\times10^{-4}$ \\
HD202917& $<2.8\,\mu$m\tablenotemark{d}	& 0.3$\,\mu$m	& {\it 1000$\,\mu$m}	&$-3.4\pm0.1	$	& $46^{+9}_{-3}$ AU		& {\it 80 AU}		& {\it-1.5}	& 0.70	& $3.4\times10^{-4}$\\
HD3003	& $3.5^{+0.5}_{-0.3}\,\mu$m& 3.4$\,\mu$m	& {\it 1000$\,\mu$m}	&$-4.4^{+0.1}_{-0.2}$	& $7.8^{+0.3}_{-0.2}$ AU	& {\it 120 AU}		& {\it-1.5}	&$8.0\times10^{-11}$	& $7.0\times10^{-6}$ \\
HD12039& {\it 0.4$\,\mu$m}			& 0.4$\,\mu$m	& {\it 1000$\,\mu$m}	&{\it -3.5}		& $14\pm3$ AU		& {\it 120 AU}		& {\it-1.5}	& 0.88	& $1.7\times10^{-4}$ \\
HD1466	& {\it 0.5$\,\mu$m}		& 0.5$\,\mu$m	& {\it 1000$\,\mu$m}	&{\it -3.5}		& $7.8^{+2.1}_{-1.8}$ AU	& {\it 120 AU}		& {\it-1.5}	& 0.30	& $6.2\times10^{-5}$ \\
\hline
\end{tabular}
\tablecomments{Model parameters in italics are fixed.}
\begin{tablenotes}
\item[a]{Model parameters}
\item[b]{Model goodness-of-fit} 
\item[c]{Lower limit on dust mass is calculated from the model parameters}
\item[d]{Unconstrained parameter, limits given are 3$\sigma$ confidence}
\end{tablenotes}
\end{threeparttable}
\end{table*}

\subsubsection{HD202917}

HD202917 is a G5V star 46 pc away.  The disk was observed with {\it Spitzer} by \citet{Bryden06} and analyzed as part of the FEPS survey by \citet{Hillenbrand08} and \citet{Carpenter08}.  \citet{Smith06} fit the disk with a blackbody grain model to get a lower limit on the disk inner radius. They found the inner radius to be $>7.4$ AU.  \citet{Hillenbrand08} fit a multi-temperature blackbody model to the disk that gave an inner radius of 2.5 AU, also a lower limit.  HD 202917 was also resolved in scattered light by HST/ACS, giving a disk  outer radius of R $\approx 80$ AU \citep{Krist07,Mustill09}.  For this reason, we fixed $r_{max} = 80$ AU rather than the usual value of 120 AU.

The fit is not well constrained in $a_{min}$, but the fit with the lowest $\chi^2$ value has $a_{min} = 0.3\,\mu$m, equal to the blowout size.  We found the best fitting inner radius to be 46 AU.  This result is larger than the previous lower limits due to the presence of small grains in the model.  We find this disk to be consistent with a belt of material between 45 and 80 AU.  

\subsubsection{HD3003}

HD3003 is an A0V star with a warm disk first detected by IRAS \citep{Oudmaijer92}.  \citet{Smith06} also observed the HD3003 disk with {\it Spitzer} MIPS at 24 and 70$\,\mu$m.  \citet{Smith06} modeled the disk with blackbody grains at a single radius and found a dust temperature of 230 K at a radius of 6.7 AU.  \citet{Smith10} added unresolved ground based mid-IR photometry from TIMMI2, VISIR, Michelle and TReCS and found a blackbody temperature of 265 K with a radius of 4 AU.  \citet{Smith10} also make the point that HD3003 is a binary, and a disk of this temperature must be circumstellar not circumbinary in order to be stable.  

With the new PACS data, we found a lower blackbody temperature of 208 K.  This is still highest temperature of all the disks in the sample, implying the grains are either very close to the star, or smaller than 1$\,\mu$m in size.  But HD3003 is an A0 star, and has a blowout size of 3.4$\,\mu$m.  Therefore, the temperature is likely due to the distance from the star.  Our model of the HD3003 disk gives an inner radius of $7.8$ AU and a minimum grain size consistent with the estimated blowout size. The goodness-of-fit is very small for HD3003, with $Q = 8\times10^{-11}$, despite reproducing the PACS data quite well.  This small value is driven mostly by the small error bars on the {\it Spitzer} IRS spectrum.  However, this goodness-of-fit is indeed larger than the value found for the blackbody models.  

The grain size distribution, however, departs from the expected shape.  The best fit grain size distribution power-law index value of $\kappa = -4.4$ is much steeper than the $\kappa=-3.5$ value expected for a steady state collisional cascade.  
HD3003 is the only binary system with a confirmed disk in the sample.  The apparent binary separation is $\sim0.1^{\prime\prime}$ \citep{Mason01}.  If this projected separation were a true binary separation, the companion would be about 4.6 AU from HD3003.  \citet{Smith10} suggest the binary must have a semi-major axis $>14.4$ AU for the disk to be stable.  An unstable system could provide one explanation for the large departure of $\kappa$ from the steady state value.  Other possibilities are explored in Section 6.3.  

\subsubsection{HD12039}

HD12039 is a G3/5V star 42 pc away with a debris disk first detected using data from the FEPS Legacy Program \citep{Hines06}.  The FEPS team analyzed {\it Spitzer} observations with IRAC, IRS, and MIPS (24, 70, and 160$\,\mu$m).  The disk was not detected with {\it Spitzer} at 70 and 160$\,\mu$m.  A blackbody fit gave a characteristic disk temperature of 110 K and a lower limit on the radius of 6 AU. \citet{Hines06} also fit the disk with the models of \citet{Wolf03} that uses astronomical silicates, with flat surface density profile ($q=0$) and a \citet{Dohnanyi69} distribution of grains from 0.4-1000$\,\mu$m.  The best fitting model yielded an inner radius of 28 AU.  

The best-fitting model in \citet{Hines06} was very dependent on the MIPS 70$\,\mu$m upper limit.  With the PACS 70$\,\mu$m detection, the disk radius is better constrained.  We fit the disk with a surface density profile of $q=-1.5$, a minimum grain size at the blowout limit, and a \citet{Dohnanyi69} grain size distribution, similar to the modeling done by \citet{Hines06}.  But, in contrast to \citet{Hines06}, we found an inner radius of $14\pm3$ AU.  This difference is due mainly to our measured disk flux at $70\,\mu$m that \citet{Hines06} did not have.    

\subsubsection{HD1466}

HD1466 is an F9V star 41 pc away with an excess detected by \citet{Smith06} at both 24 and 70$\,\mu$m with {\it Spitzer} MIPS.  
\citet{Smith06} found a minimum radius of 7.2 AU. 
Our physical disk model fit gave an inner radius of 7.8 AU, consistent with previous results. 

\subsubsection{HD30051}

The HD30051 disk was first discovered in this survey.  Unfortunately, the disk system was never observed with {\it Spitzer}, and very little mid-IR data are available to constrain the disk parameters.  With only two data points showing IR excess, we were unable to model the disk in any more detail than a blackbody model.  But the best-fitting blackbody temperature of 58 K indicates the main component of the disk is far from the star.  Keeping the assumption of large pure blackbody grains, the disk would be centered around $\sim 45$ AU.  This is a lower limit on the disk radius because smaller silicate grains would have a temperature of 58 K tens of AU farther out.  

\begin{figure*}
\plottwo{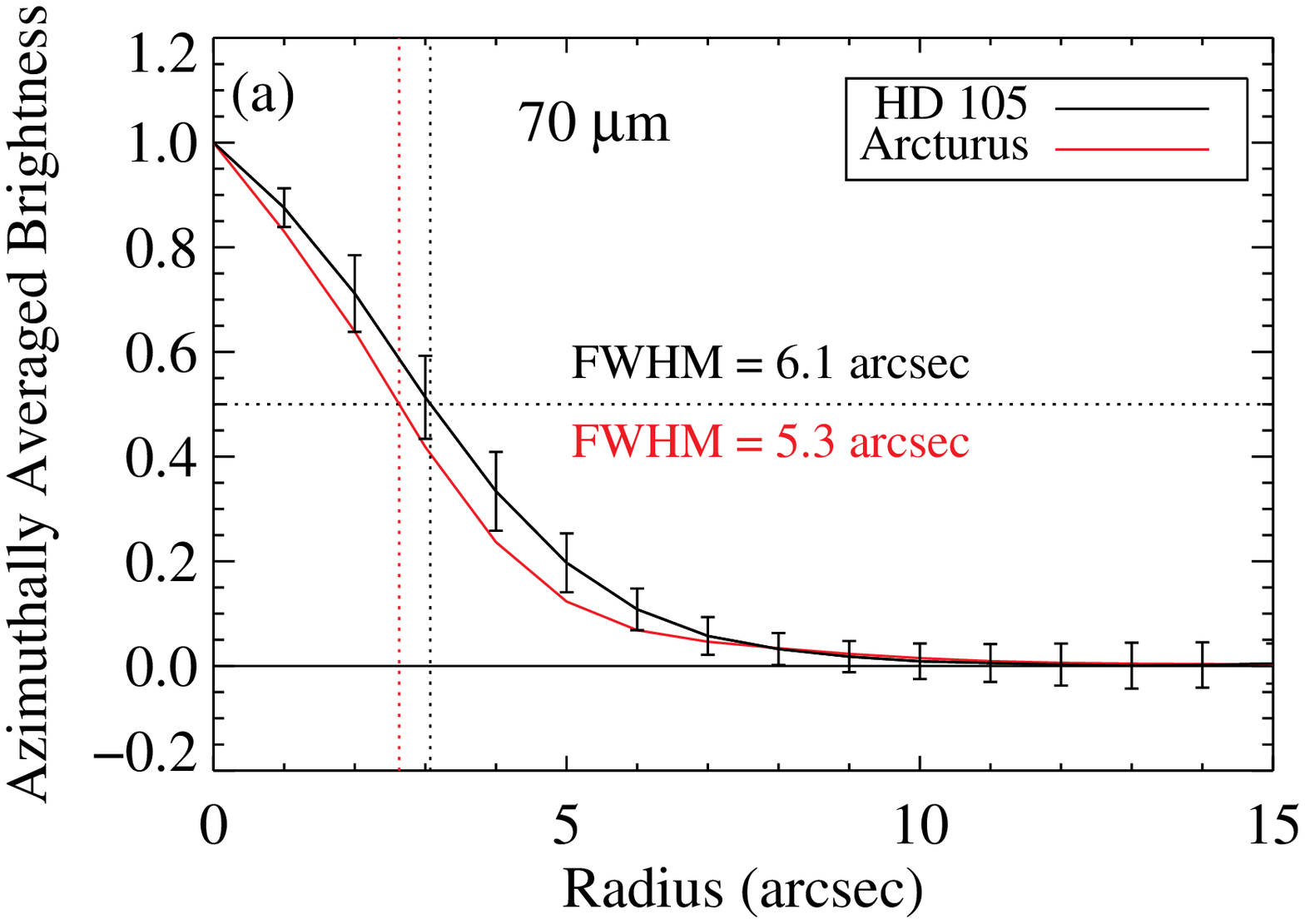}{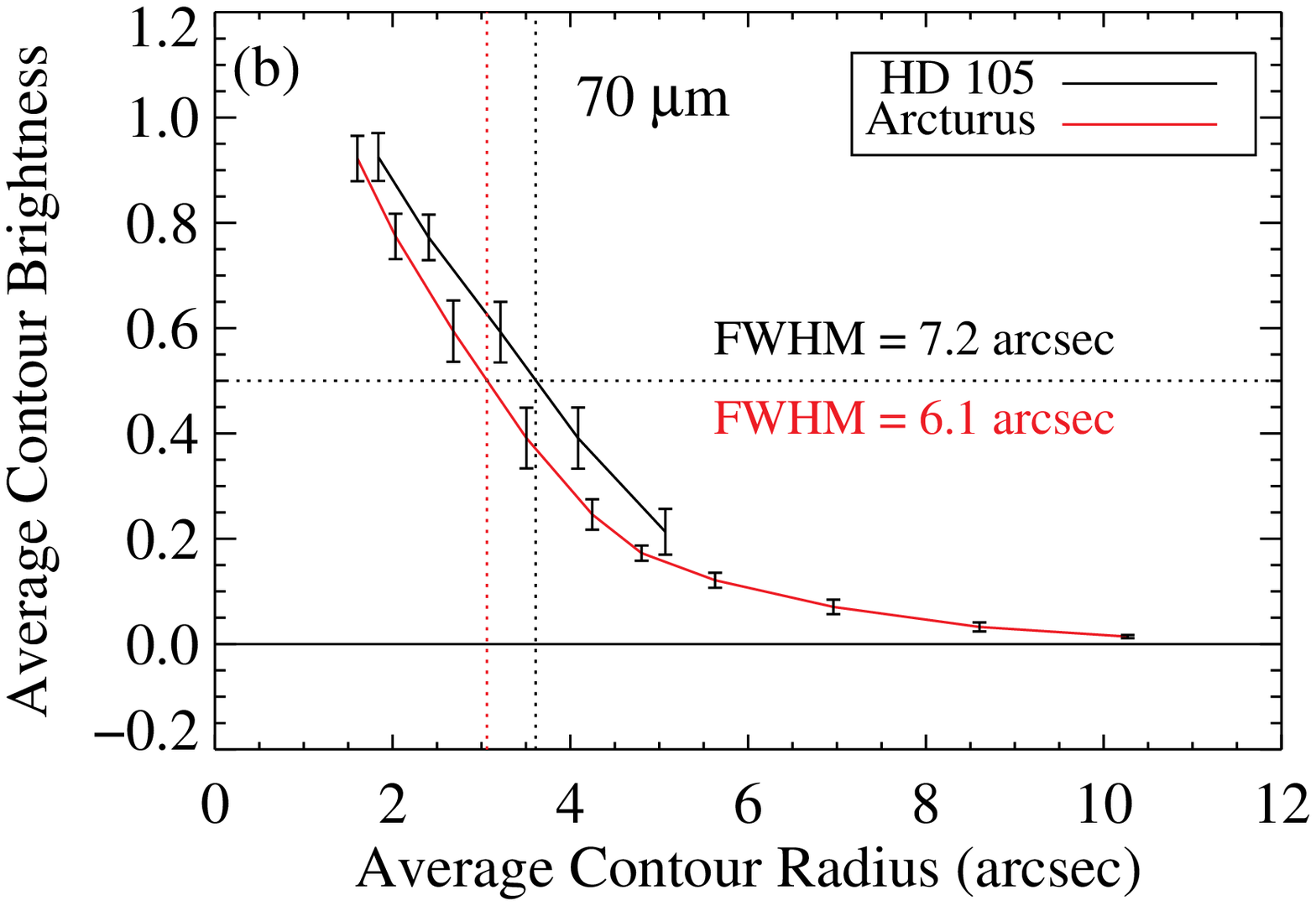}
\caption{(a): Azimuthally averaged radial brightness profiles for HD105 and the reference star Arcturus at 70$\,\mu$m.  The radial profiles are azimuthally averaged over annuli with one arcsecond widths. (b): Averaged brightness profiles from contours for HD105 and Arcturus at 70$\,\mu$m.  HD105 is extended beyond the PSF in both profiles. \label{fig:radprof}}
\end{figure*}

\section{Resolving the HD105 Debris Disk\label{resolved}}

\subsection{Radial Profile}

\begin{figure}
\plotone{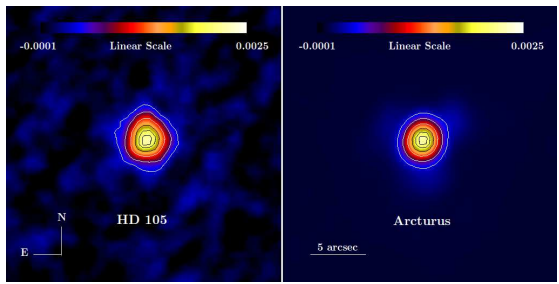}
\caption{Images of HD105 (left) and a PSF reference star Arcturus (right) at 70$\,\mu$m.  Brightness contours are overlaid at 15, 30, 50, 70, 85 and 95\% of the peak brightness. \label{fig:contours}}
\end{figure}

One disk in our sample, HD105, is marginally resolved at 70$\,\mu$m in the {\it Herschel} PACS images.  HD105 is the brightest disk in our sample.  It may not be the most extended, but for the other disks an extended structure would fall below the background noise.  We determined that the HD105 disk was resolved in two ways.  The first was to compare the azimuthally averaged radial profile of the disk to a reference star, Arcturus.  Figure \ref{fig:radprof} (a) shows the azimuthally averaged radial profiles of the two images.  These profiles were calculated by measuring the mean flux within annuli of one arcsecond width around the star.  Scan maps with a smaller pixel scale of $1^{\prime\prime}$ pixel$^{-1}$ were created for this purpose. The uncertainties in the measurements were determined by calculating the standard deviation in the flux within each annulus.  This method leads to an overestimation of the errors because the PACS PSF changes radially across each annulus due to its tri-lobed shape. The azimuthally averaged radial profile of HD105 shown in Figure \ref{fig:radprof} (a) is extended beyond the PSF with a FWHM of $6.1\pm0.8^{\prime\prime}$ compared to the PSF FWHM of $5.3\pm0.5^{\prime\prime}$.  The error in the FWHM is determined from the uncertainties in the brightness by calculating the distance away from the FWHM where the brightness profile plus and minus the errors would be equal to half the maximum brightness.  This is not a strong detection of extended structure given the large (and overestimated) errors.  Therefore, we also used a second method to determine if the disk is resolved.  

The second method tries to avoid problems with the PACS PSF tri-lobe shape by averaging over brightness contours rather than annuli.  Figure \ref{fig:contours} shows the PACS 70$\,\mu$m images of HD105 and the reference star Arcturus scaled to the same peak brightness with brightness contours overlaid. The fluxes in the regions between the contours were averaged and the uncertainty measured from the standard deviation.  The contours used for Figure \ref{fig:radprof} (b) are 15, 30, 50, 70, 85, and 100\% of the peak brightness for HD105 and Arcturus. For Arcturus, which is brighter than HD105, the 1, 2, 5, and 10\% contours are also used. The radius adopted for each mean flux value was the mean distance of the region between the contours.  This method is unable to map the brightness profile of HD105 far from the star because it quickly falls below the background.  But it it clear from Figure \ref{fig:radprof} (b) that the disk is extended with a FWHM of $7.2\pm0.5^{\prime\prime}$ beyond the PSF that has a FWHM of $6.1\pm0.4^{\prime\prime}$.

The PACS PSF size is given by the {\it Herschel} PACS ICC as a two-dimensional Gaussian with FWHM of $5.5\times5.8^{\prime\prime}$.  The FWHM reported here for HD105 was not calculated in the same way.  Therefore, the FWHM values should not be directly compared to the PACS PSF FWHM when determining the physical size of the disk.  The FWHM values are given only to show that the disk is indeed significantly extended beyond the PSF.  In Section 5.2, we determine the physical size of the disk without using the FWHM.

\subsection{Determining the Outer Radius}

Without data at wavelengths longer than 160$\,\mu$m, the outer radius cannot be determined from SED fitting alone.  However, resolved imaging provides geometrical information.  The marginally resolved image of HD105 constrains the outer radius of the disk.  

To determine the outer radius of the disk, we produced a synthetic image of the disk using the parameters from the best fitting SED model.  We then convolved our model with the PACS PSF\footnotemark[14] and rotated it to the position angle of the observation for direct comparison.
We varied the outer radius and determined the $\chi^2$ value between the model and the images of HD105.  
Once the outer radius of the model is greater than a certain value, the surface brightness at that radius is so low that it falls below the noise level.  Therefore, we can only place a lower limit on the outer radius of the disk.  We found a disk outer radius of $>280$ AU.

\footnotetext[14]{\url{http://pacs.ster.kuleuven.ac.be/pubtool/PSF/}}

There are, of course, several assumptions that go into this model.  The first is that the ring is circular and face-on.  The second is that the dust surface density distribution follows an r$^{-1.5}$ profile.  This is expected to be the case for collisionally dominated disks.  Several resolved images of disks have shown more complicated structure such as rings with sharper density profiles, clumps, warps, and other asymmetries.  These structures would have less of an effect on the SED, which is fairly insensitive to the density profile, but would strongly affect the analysis of the {\it Herschel} images.  The outer radius derived here is the first such constraint put on the disk, but an image with higher spatial resolution is needed to better determine the spatial extent. 

\section{Discussion\label{discuss}}

We detected IR-excesses in $\sim 1/3$ of the stars in our sample.  The Tucana-Horologium disks we detect all have some similarities.  They are optically thin debris disks with relatively low fractional luminosities compared with younger protoplanetary disks, which have typical fractional luminosities of L$_{\text{IR}}$/L$_\ast \sim 0.1$ \citep{Cieza12}.  As of yet, none of the Tucana-Horologium systems have shown any significant amount of gas.  However, these disks do show a remarkable amount of variety for systems of the same age ($\sim30$ Myrs).  The single temperature blackbody fits give a range in temperature of 50-208 K.  Although the hottest disk is around an A0 star (HD3003), the other six disks still display a temperature range of 50-146 K with no apparent dependence on spectral type. Any trends that may exist would be hard to see in a such a small sample. Our further modeling also shows a variety in other disk properties.  Three of these properties, minimum grain size, the grain size distribution power-law, and inner holes, will be discussed in the following sections.

\subsection{Minimum Grain Size}

The minimum grain size in debris disks is expected to be approximately equal to the blowout size due to radiation pressure.  Any grains smaller than this limit would be ejected from the system on a timescale of a few thousand years.  But there are a few problems with using the blowout size as a limit.  First, there are many assumptions that go into the blowout size calculation.  The grains are assumed to be spherical, have a constant density, and start out on circular orbits.  Additionally, the grain composition must be assumed.  As circumstellar grain composition is hard to determine due to the lack of mid-IR solid state features from most debris disks, \citep[e.g.\ ][]{Jura04}, ISM grain composition is commonly used.  But densities can differ greatly between silicates, graphite and ices, all of which are expected to be present in debris disks.  Porosity may also play a role in changing this grain size limit \citep{Lebreton12}.  For these reasons, it is difficult to interpret the minimum grain size results based on the blowout size.  

Only two disks in our sample had well constrained minimum grain sizes.  For HD105, the minimum grain size found was more than an order or magnitude greater than the blowout size.  Here, either the calculated blowout size is inaccurate, pointing to a different grain composition or porosity, or larger grains are efficiently removed by some other mechanism.  \citet{Thebault08} show that dynamically cold disks have fewer small grains.  The smaller velocity of the collisions decreases the production rate of the small grains while the destruction rate increases.  This produces a disk dominated by larger grains with orbits mostly confined to the planetesimal ring, very similar to what is observed around HD105.  For HD3003, the minimum grain size is similar to the blowout size.

\subsection{Inner Holes}

The range of temperatures seen in the blackbody fits imply inner gaps with radii from 4.5 to 52 AU.  The largest inner hole is in the HD105 disk. 
Giant planets could be responsible for these holes.  However, there are also viable mechanisms to explain the holes without planets \citep{Kennedy10}.  For instance, dust parent bodies may preferentially form at outer locations.  This can be expected as a result of photoevaporative clearing of the inner gas disk and pile-up of dust at its inner edge \citep[e.g.][]{Alexander07} or as a consequence of rapid planetesimal formation in spiral arms of a self-gravitating disk \citep[e.g.][]{Rice06}.
Resolved imaging showing the sharpness of the hole's edge could put more constraints on the processes that created the inner hole, as was done in the case of the Fomalhaut debris ring \citep{Chiang09}.

\subsection{An Unusual Debris Disk?}
The disk of HD3003 was found to have a grain size distribution much steeper than the typical \citet{Dohnanyi69} steady state collisional cascade.  A steeper distribution implies two possibilities: the disk has an overabundance of small grains, or a paucity of large grains.  The minimum grain size we find for HD3003 ($a_{min} = 3.5\,\mu$m) is large enough that we suspect the explanation to be the paucity of large grains.  We believe the minimum grain size to be correct, as it is consistent with both the blowout size and the lack of a $10\,\mu$m silicate feature in the IRS spectrum which is only seen in disks with small (sub-$\mu$m) grains \citep{Kessler-Silacci06}.

We investigated other explanations of the SED slope.  The lack of flux at longer wavelengths could mean a smaller outer radius than was assumed.  The disk could be tidally truncated by its binary companion if the other star were close enough to the disk.  We tried fixing the grain size distribution power-law index to $\kappa =-3.5$ and varying $r_{max}$ from 15-120 AU.  These models were not able to reproduce the observed data.  

The unusual behavior of the HD3003 SED has been seen in older debris disks in the DUNES sample \citep{Eiroa10}.  \citet{Ertel12} detected three debris disks with {\it Herschel} whose slopes are also inconsistent with a \citet{Dohnanyi69} distribution.  They give three possible explanations for the underabundance of large grains.  First, there is a departure from the steady state collisional cascade conditions.  HD3003 is a denser disk, so it should be collision dominated not transport dominated.  However, the distribution can still deviate from \citet{Dohnanyi69} by processes such as radiation pressure that causes a wavy distribution that is steeper at some points \citep{CampoBagatin94,Thebault03}.  The second possibility is that grains of a different composition would have a different absorption coefficient, $Q_{abs}$, which may affect the SED slope.  And lastly, a sheparding planet could result in a spatial separation of the small and large grains, leaving the large grains undetectable. 

An idea not proposed in \citet{Ertel12}, but possibly relevant here, is enhanced stirring of the planetesimal disk by the companion star.  The true separation of HD3003's companion star is unknown.  If it is close enough, it could violently stir the disk.  If the orbit of the binary is eccentric, then a close passage of the companion would excite the disk, raising the mean eccentricity of the disk particles, and thereby increasing their relative velocities \citep{Mustill09}.  Numerical simulations show that the waviness of the grain size distribution depends on the collision velocities.  The amplitude and the peak-to-peak wavelength of ripples in the grain size distribution increase with higher velocities \citep{Thebault03,Krivov06,Wyatt11}.  Larger ripples mean the size distribution will be steeper for grain sizes above the blowout limit, as is seen in HD3003.
However, the strange behavior of the SED is not yet understood, and will require more data and further modeling to determine its cause.  

\section{Summary/Conclusion}
We observed seventeen stars in the Tucana-Horologium Association with the PACS instrument on the {\it Herschel Space Observatory}.  We detected six debris disks, including one previously unknown disk and put sensitive upper limits on those not detected.  We modeled the disks with a thermal dust disk model and were able to place tighter constraints on several disk parameters, such as the inner disk radius, minimum grain size, and grain size distribution.  Additionally, we marginally resolved one disk and were able to put a lower limit on the outer radius.  Future work will include {\it Herschel} SPIRE observations to better populate the sub-mm portion of the SEDs, and resolved imaging with ALMA to break degeneracies by determining the disk geometry.  These data will also be combined with other targets of different ages to examine the statistical properties of the entire GASPS sample.  

\acknowledgements
This work is based on observations made with Herschel, a European Space Agency Cornerstone Mission with significant participation by NASA. Support for this work was provided by NASA through an award issued by JPL/Caltech. JCA thanks the PNP-CNES and the French National Research Agency (ANR) for financial support through contract ANR-2010 BLAN-0505-01 (EXOZODI).


\begin{center}
\begin{longtable}{l c c l}
\caption[]{Archive Data 
Used In SED Fitting.} \label{tab:alldata} \\

\hline \multicolumn{4}{c}{HD105} \\ \hline 
\endfirsthead

\multicolumn{4}{c}%
{{\bfseries \tablename\ \thetable{} -- continued from previous page}} \\
\hline \multicolumn{1}{l}{Instrument} &
\multicolumn{1}{c}{Wavelength} &
\multicolumn{1}{c}{Flux} & \multicolumn{1}{l}{Reference} \\ 
\multicolumn{1}{}{ } &
\multicolumn{1}{c}{($\mu$m)} &
\multicolumn{1}{c}{(mJy)} & \multicolumn{1}{}{  } \\ \hline 
\endhead

\hline \multicolumn{3}{r}{{Continued on next page}} \\ \hline
\endfoot

\hline \hline
\endlastfoot
System  & Wavelength & Flux & Reference\\
	    & ($\mu$m)	& (mJy) & 	\\
\hline
Hipparcos 	& 0.44	& $2020\pm27.91$	&\cite{Hog00} \\
Hipparcos 	& 0.55	& $3504\pm32.27$	&\cite{Hog00} \\
2MASS		& 1.25	& $4139\pm76.25$	&\cite{Cutri03} \\
2MASS		& 1.65	& $3425\pm69.41$	&\cite{Cutri03} \\
2MASS		& 2.17	& $2383\pm43.90$	&\cite{Cutri03} \\
{\it Spitzer}/IRAC	& 3.6 	& $1023\pm7.36$		&\cite{Carpenter08}\\
{\it Spitzer}/IRAC	& 4.5 	& $645.4\pm7.87$	&\cite{Carpenter08}\\
{\it Spitzer}/IRAC	& 5.8 	& $410.5\pm4.27$	&\cite{Carpenter08}\\
{\it Spitzer}/IRAC	& 8.0 	& $230.7\pm1.52$	&\cite{Carpenter08}\\
AKARI		& 9	& $223.1\pm9.59$	&\cite{Yamamura10}\\
{\it Spitzer}/IRS	&5-37&		&\cite{Carpenter08}\\
{\it Spitzer}/MIPS	& 24	& $28.29\pm0.25$	&\cite{Carpenter08}\\
\hline
\multicolumn{4}{c}{HD202917}\\
\hline
System  & Wavelength & Flux & Reference\\
	    & ($\mu$m)	& (mJy) & 	\\
\hline
Hipparcos 	& 0.44	& $619.2\pm10.84$	& \cite{Hog00} \\
Hipparcos 	& 0.55	& $1233\pm14.76$	& \cite{Hog00} \\
2MASS		& 1.25	& $1770\pm34.25$	&\cite{Cutri03} \\
2MASS		& 1.65	& $1585\pm55.47$	&\cite{Cutri03} \\
2MASS		& 2.17	& $1150\pm23.31$	&\cite{Cutri03} \\
WISE 		& 3.4	& $575.86\pm17.23$	&\cite{Wright10}	\\
{\it Spitzer}/IRAC	& 3.6 	& $519.2\pm3.74$	&\cite{Carpenter08}\\
{\it Spitzer}/IRAC	& 4.5 	& $320.8\pm3.91$	&\cite{Carpenter08}\\
WISE 		& 4.6	& $304.36\pm5.94$	&\cite{Wright10}	\\
{\it Spitzer}/IRAC	& 8.0 	& $117.3\pm1.44$	&\cite{Carpenter08}\\
AKARI		& 9	& $96.55\pm12.4$	&\cite{Yamamura10}\\
IRAS  		& 12	& $101.5\pm23.8$	&\cite{Moshir92}\\
WISE 		& 12	& $58.93\pm1.04$	&\cite{Wright10}	\\
WISE 		& 22	& $19.82\pm1.01$	&\cite{Wright10}	\\
{\it Spitzer}/IRS	&5-37&		&\cite{Carpenter08}\\
{\it Spitzer}/MIPS	& 24	& $20\pm0.8$		&\cite{Smith06} \\
\hline
\multicolumn{4}{c}{HD3003}\\
\hline
Instrument  & Wavelength & Flux & Reference\\
	    & ($\mu$m)	& (mJy) & 	\\
\hline
Hipparcos 	& 0.44	& $35400\pm456.5$	& \cite{Hog00} \\
Hipparcos 	& 0.55	& $35290\pm292.6$	& \cite{Hog00} \\
2MASS		& 1.25	& $15070\pm513.6$	&\cite{Cutri03} \\
2MASS		& 1.65	& $8870\pm621.4$	&\cite{Cutri03} \\
2MASS		& 2.17	& $6760\pm124.5$	&\cite{Cutri03} \\
AKARI		& 9	& $586.1\pm9.77$	&\cite{Yamamura10}\\
IRAS  		& 12	& $446.0\pm31.22$	&\cite{Moshir92}\\
AKARI		& 18	& $237.7\pm15.7$	&\cite{Yamamura10}\\
{\it Spitzer}/IRS	&5-37&		&This Work\\
{\it Spitzer}/MIPS	& 24	& $223.9\pm9.0$		&\cite{Smith06} \\
\hline
\multicolumn{4}{c}{HD12039}\\
\hline
Instrument  & Wavelength & Flux & Reference\\
	    & ($\mu$m)	& (mJy) & 	\\
\hline
Hipparcos 	& 0.44	& $1103\pm17.27$	& \cite{Hog00} \\
Hipparcos 	& 0.55	& $2073\pm22.91$	& \cite{Hog00} \\
2MASS		& 1.25	& $2885\pm61.11$	&\cite{Cutri03} \\
2MASS		& 1.65	& $2445\pm85.59$	&\cite{Cutri03} \\
2MASS		& 2.17	& $1718\pm41.15$	&\cite{Cutri03} \\
{\it Spitzer}/IRAC	& 3.6 	& $747.3\pm5.38$	&\cite{Carpenter08}\\
{\it Spitzer}/IRAC	& 4.5 	& $470.9\pm5.75$	&\cite{Carpenter08}\\
{\it Spitzer}/IRAC	& 8.0 	& $170.4\pm1.13$	&\cite{Carpenter08}\\
AKARI		& 9	& $159.9\pm6.9$		&\cite{Yamamura10}\\
{\it Spitzer}/IRS	&7-37&		&\cite{Carpenter08}\\
{\it Spitzer}/MIPS	& 24	& $25.65\pm0.23$	&\cite{Carpenter08} \\
\hline
\multicolumn{4}{c}{HD1466}\\
\hline
Instrument  & Wavelength & Flux & Reference\\
	    & ($\mu$m)	& (mJy) & 	\\
\hline
Hipparcos 	& 0.44	& $2248\pm31.06$	&\cite{Hog00} \\
Hipparcos 	& 0.55	& $3686\pm37.34$	&\cite{Hog00} \\
2MASS		& 1.25	& $4147\pm68.75$	&\cite{Cutri03} \\
2MASS		& 1.65	& $3244\pm107.6$	&\cite{Cutri03} \\
2MASS		& 2.17	& $2314\pm36.23$	&\cite{Cutri03} \\
AKARI		& 9	& $207.1\pm20.8$	&\cite{Yamamura10}\\
IRAS 		& 12	& $141\pm16.92$		&\cite{Moshir92}\\
{\it Spitzer}/IRS	&5-37&		&This Work\\
{\it Spitzer}/MIPS	& 24	& $32.90\pm1.3$		&\cite{Smith06}\\
\hline
\multicolumn{4}{c}{HD30051}\\
\hline
Instrument  & Wavelength & Flux & Reference\\
	    & ($\mu$m)	& (mJy) & 	\\
\hline
Hipparcos 	& 0.44	& $3660\pm50.56$	&\cite{Hog00} \\
Hipparcos 	& 0.55	& $5163\pm47.55$	&\cite{Hog00} \\
2MASS		& 1.25	& $4867\pm89.66$	&\cite{Cutri03} \\
2MASS		& 1.65	& $3829\pm119.9$	&\cite{Cutri03} \\
2MASS		& 2.17	& $2599\pm52.66$	&\cite{Cutri03} \\
WISE 		& 3.4	& $1329\pm61.35$	&\cite{Wright10}\\
WISE 		& 4.6	& $717.4\pm14.69$	&\cite{Wright10}\\
AKARI		& 9	& $237.4\pm21.9$	&\cite{Yamamura10}\\
IRAS 		& 12	& $152\pm15.2$		&\cite{Moshir92}\\
WISE 		& 12	& $126.2\pm2.11$	&\cite{Wright10}\\
WISE 		& 22	& $34.67\pm1.37$	&\cite{Wright10}\\
\hline
\multicolumn{4}{c}{HD16978}\\
\hline
Instrument  & Wavelength & Flux & Reference\\
	    & ($\mu$m)	& (mJy) & 	\\
\hline
Hipparcos 	& 0.44	& $95730\pm1234$	&\cite{Hog00} \\
Hipparcos 	& 0.55	& $87120\pm722.1$	&\cite{Hog00} \\
2MASS		& 1.25	& $26620\pm7349$	&\cite{Cutri03} \\
2MASS		& 1.65	& $17260\pm4337$	&\cite{Cutri03} \\
2MASS		& 2.17	& $13250\pm439.5$	&\cite{Cutri03} \\
AKARI		& 9	& $1068\pm6.11$		&\cite{Yamamura10}\\
IRAS 		& 12	& $743.0\pm66.87$	&\cite{Moshir92}\\
AKARI		& 18	& $257.7\pm12.1$	&\cite{Yamamura10}\\
{\it Spitzer}/MIPS	& 24	& $124.0\pm4.96$	&\cite{Rebull08}\\
\hline
\multicolumn{4}{c}{HD2884}\\
\hline
Instrument  & Wavelength & Flux & Reference\\
	    & ($\mu$m)	& (mJy) & 	\\
\hline
Hipparcos 	& 0.44	& $79260\pm1022$	&\cite{Hog00} \\
Hipparcos 	& 0.55	& $72860\pm604$		&\cite{Hog00} \\
2MASS		& 1.25	& $21720\pm5128$	&\cite{Cutri03} \\
2MASS		& 1.65	& $13790\pm965.9$	&\cite{Cutri03} \\
2MASS		& 2.17	& $10750\pm356.6$	&\cite{Cutri03} \\
AKARI		& 9	& $895.2\pm31.0$	&\cite{Yamamura10}\\
IRAS 		& 12	& $1300\pm78$		&\cite{Moshir92}\\
AKARI		& 18	& $188.4\pm14.8$	&\cite{Yamamura10}\\
{\it Spitzer}/IRS	&5-37&		&This work\\
{\it Spitzer}/MIPS	& 24	& $101.7\pm8.6$		& \cite{Smith06}\\
\hline
\multicolumn{4}{c}{HD224392}\\
\hline
Instrument  & Wavelength & Flux & Reference\\
	    & ($\mu$m)	& (mJy) & 	\\
\hline
Hipparcos 	& 0.44	& $37070\pm478$		&\cite{Hog00} \\
Hipparcos 	& 0.55	& $37710\pm312.6$	&\cite{Hog00} \\
2MASS		& 1.25	& $17320\pm590.3$	&\cite{Cutri03} \\
2MASS		& 1.65	& $10730\pm306.5$	&\cite{Cutri03} \\
2MASS		& 2.17	& $7840\pm151.7$	&\cite{Cutri03} \\
AKARI		& 9	& $660.5\pm18.2$	&\cite{Yamamura10}\\
IRAS 		& 12	& $477\pm28.62	$	&\cite{Moshir92}\\
AKARI		& 18	& $113\pm15.59$		&\cite{Yamamura10}\\
\hline
\multicolumn{4}{c}{HD2885}\\
\hline
Instrument  & Wavelength & Flux & Reference\\
	    & ($\mu$m)	& (mJy) & 	\\
\hline
Hipparcos 	& 0.44	& $54530\pm703.1$	&\cite{Hog00} \\
Hipparcos 	& 0.55	& $59330\pm491.8$	&\cite{Hog00} \\
2MASS		& 1.25	& $29820\pm8003$	&\cite{Cutri03} \\
2MASS		& 1.65	& $21340\pm4192$	&\cite{Cutri03} \\
2MASS		& 2.17	& $15160\pm502.8$	&\cite{Cutri03} \\
AKARI		& 9	& $1291\pm18.6$		&\cite{Yamamura10}\\
AKARI		& 18	& $340.6\pm25.6$	&\cite{Yamamura10}\\
{\it Spitzer}/MIPS	& 24	& $156.1\pm4.5$		&\cite{Smith06}\\
\hline
\multicolumn{4}{c}{HD53842}\\
\hline
Instrument  & Wavelength & Flux & Reference\\
	    & ($\mu$m)	& (mJy) & 	\\
\hline
Hipparcos 	& 0.44	& $2529\pm37.28$	&\cite{Hog00} \\
Hipparcos 	& 0.55	& $3716\pm37.65$	&\cite{Hog00} \\
2MASS		& 1.25	& $3813\pm101.9$	&\cite{Cutri03} \\
2MASS		& 1.65	& $2831\pm80.83$	&\cite{Cutri03} \\
2MASS		& 2.17	& $2015\pm38.98$	&\cite{Cutri03} \\
WISE 		& 3.4	& $894.4\pm31.9$	&\cite{Wright10}\\
WISE 		& 4.6	& $525.0\pm10.7$	&\cite{Wright10}\\
AKARI		& 9	& $179.4\pm8.49$	&\cite{Yamamura10}\\
IRAS 		& 12	& $140.0\pm10.0$	&\cite{Moshir92}\\
WISE 		& 12	& $100.1\pm16.7$	&\cite{Wright10}\\
WISE 		& 22	& $40.4\pm1.29$		&\cite{Wright10}\\
{\it Spitzer}/IRS	&7-37&		&\cite{Moor09}\\
{\it Spitzer}/MIPS	& 24	& $31.0\pm1.3$		&\cite{Moor09}\\
\hline
\multicolumn{4}{c}{HD44627}\\
\hline
Instrument  & Wavelength & Flux & Reference\\
	    & ($\mu$m)	& (mJy) & 	\\
\hline
Hipparcos 	& 0.44	& $318.1\pm8.21$	&\cite{Hog00} \\
Hipparcos 	& 0.55	& $767.2\pm12.72$	&\cite{Hog00} \\
2MASS		& 1.25	& $1486\pm32.86$	&\cite{Cutri03} \\
2MASS		& 1.65	& $1497\pm28.95$	&\cite{Cutri03} \\
2MASS		& 2.17	& $1075\pm23.77$	&\cite{Cutri03} \\
WISE 		& 3.4	& $533.0\pm14.93$	&\cite{Wright10}		\\
WISE 		& 4.6	& $295.5\pm5.77$	&\cite{Wright10}		\\
AKARI		& 9	& $181\pm5.09$		&\cite{Yamamura10}\\
IRAS 		& 12	& $101\pm16.16$		&\cite{Moshir92}\\
WISE 		& 12	& $58.39\pm1.03$	&\cite{Wright10}		\\
WISE 		& 22	& $15.06\pm0.81$	&\cite{Wright10}		\\
\hline
\multicolumn{4}{c}{HD55279}\\
\hline
Instrument  & Wavelength & Flux & Reference\\
	    & ($\mu$m)	& (mJy) & 	\\
\hline
Hipparcos 	& 0.44	& $110.5\pm6.01$	&\cite{Hog00} \\
Hipparcos 	& 0.55	& $304.0\pm8.96$	&\cite{Hog00} \\
2MASS		& 1.25	& $787.9\pm16.69$	&\cite{Cutri03} \\
2MASS		& 1.65	& $754.9\pm39.65$	&\cite{Cutri03} \\
2MASS		& 2.17	& $579.6\pm13.88$	&\cite{Cutri03} \\
WISE 		& 3.4	& $273.9\pm6.38$	&\cite{Wright10}		\\
WISE 		& 4.6	& $148.1\pm2.89$	&\cite{Wright10}		\\
AKARI		& 9	& $65.96\pm5.85$	&\cite{Yamamura10}\\
WISE 		& 12	& $29.78\pm0.61$	&\cite{Wright10}		\\
WISE 		& 22	& $8.36\pm1.23$		&\cite{Wright10}		\\
\hline
\multicolumn{4}{c}{HD3221}\\
\hline
Instrument  & Wavelength & Flux & Reference\\
	    & ($\mu$m)	& (mJy) & 	\\
\hline
Hipparcos 	& 0.44	& $141.1\pm6.239$	&\cite{Hog00} \\
Hipparcos 	& 0.55	& $512.0\pm9.904$	&\cite{Hog00} \\
2MASS		& 1.25	& $1852\pm30.71$	&\cite{Cutri03} \\
2MASS		& 1.65	& $2098\pm65.72$	&\cite{Cutri03} \\
2MASS		& 2.17	& $1625\pm26.93$	&\cite{Cutri03} \\
AKARI		& 9	& $162.7\pm11.6$	&\cite{Yamamura10}\\
IRAS 		& 12	& $121.0\pm20.57$	&\cite{Moshir92}\\
\hline
\multicolumn{4}{c}{HIP107345}\\
\hline
Instrument  & Wavelength & Flux & Reference\\
	    & ($\mu$m)	& (mJy) & 	\\
\hline
Hipparcos 	& 0.44	& $16.24\pm4.654$	&\cite{Hog00} \\
Hipparcos 	& 0.55	& $78.94\pm8.231$	&\cite{Hog00} \\
2MASS		& 1.25	& $503.6\pm11.60$	&\cite{Cutri03} \\
2MASS		& 1.65	& $596.3\pm12.08$	&\cite{Cutri03} \\
2MASS		& 2.17	& $472.4\pm11.31$	&\cite{Cutri03} \\
WISE 		& 3.4	& $236.5\pm5.51$	&\cite{Wright10}		\\
WISE 		& 4.6	& $135.1\pm2.64$	&\cite{Wright10}		\\
WISE 		& 12	& $27.71\pm0.54$	&\cite{Wright10}		\\
WISE 		& 22	& $7.88\pm0.89$	&	\cite{Wright10}	\\
\hline
\multicolumn{4}{c}{HIP3556}\\
\hline
Instrument  & Wavelength & Flux & Reference\\
	    & ($\mu$m)	& (mJy) & 	\\
\hline
Hipparcos 	& 0.44	& $17.48\pm4.012$	&\cite{Hog00} \\
Hipparcos 	& 0.55	& $41.43\pm5.780$	&\cite{Hog00} \\
2MASS		& 1.25	& $645.8\pm11.90$	&\cite{Cutri03} \\
2MASS		& 1.65	& $730.3\pm16.14$	&\cite{Cutri03} \\
2MASS		& 2.17	& $595.3\pm14.81$	&\cite{Cutri03} \\
AKARI		& 9	& $118.0\pm14.8$	&\cite{Yamamura10}\\
{\it Spitzer}/MIPS	& 24	& $8.4\pm0.34$		&\cite{Rebull08}\\
\hline
\multicolumn{4}{c}{GSC8056-482}\\
\hline
Instrument  & Wavelength & Flux & Reference\\
	    & ($\mu$m)	& (mJy) & 	\\
\hline
Zeiss/FOTRAP 	& 0.44	& $15.13\pm4.82$	& \cite{Torres06}\\
Zeiss/FOTRAP 	& 0.55	& $54.15\pm9.77$	& \cite{Torres06}\\
2MASS		& 1.25	& $683\pm14.62$	& 	\cite{Cutri03}\\
2MASS		& 1.65	& $809.66\pm17.33$	& \cite{Cutri03}\\
2MASS		& 2.17	& $666.19\pm16.77$	& \cite{Cutri03}\\
AKARI		& 9	& $86.52\pm14.8$	&\cite{Yamamura10}\\
{\it Spitzer}/MIPS	& 24	& $9.0\pm0.36$		&\cite{Rebull08} \\

\end{longtable}
\end{center}

\end{document}